\documentstyle[aps,preprint,epsf]{revtex}
\tightenlines
\draft
\newcommand{\beq}{\begin{equation}}
\newcommand{\eeq}{\end{equation}}
\begin{document}

\title{NEW APPROACH TO DENSITY FUNCTIONAL THEORY
AND DESCRIPTION OF SPECTRA OF FINITE ELECTRON
SYSTEMS}

\author{M.Ya. Amusia$^{a,b}$,
A.Z. Msezane$^c$, and V.R. Shaginyan$^{c,d}$
\footnote{E-mail: vrshag@thd.pnpi.spb.ru}}

\address{$^{a\,}$The
Racah Institute of Physics, the Hebrew University, Jerusalem 91904,
Israel;\\ $^{b\,}$Physical-Technical Institute,
194021 St. Petersburg, Russia;\\ $^c\,$CTSPS, Clark Atlanta
University, Atlanta, Georgia 30314, USA;\\ $^{d\,}$Petersburg Nuclear
Physics Institute, Gatchina, 188300, Russia}

\maketitle

\begin{abstract}
The self consistent version of the density
functional theory is presented, which allows to calculate
the ground state and dynamic properties of finite multi-electron
systems. An exact functional equation for the effective interaction, from which one
can construct the action functional, density functional, the
response functions, and excitation spectra of the considered systems,
is outlined. In the context of the density functional theory we
consider the single particle excitation spectra of electron systems and relate the single particle spectrum to the
eigenvalues of the corresponding Kohn-Sham equations.  We find that
the single particle spectrum coincides neither with the eigenvalues of the Kohn-Sham equations nor with those of the
Hartree-Fock equations.
\end{abstract}


\section{Introduction}

The density functional theory (DFT), that originated from the
pioneering work of Hohenberg and Kohn\cite{wks}, has been extremely
effective in describing the ground state of finite many- electron
systems. Such a success gave birth to many papers concerned with the
generalization of DFT, which would permit the description of the excitation spectra also. The
generalization, on theoretical grounds, originated mainly
from the Runge-Gross theorem, which helped to transform DFT into
the time-dependent density functional theory TDDFT \cite{rg}.
Both, DFT and TDDFT, are based on the one-to-one correspondence
between particle densities of the considered systems and external
potentials acting upon these particles. Unfortunately, the
one-to-one correspondence establishes only the existence of the
functionals in principle, leaving aside a very important question
on how one can construct them in reality. This is why the
successes of DFT and TDDFT strongly depend upon the availability
of good approximations for the functionals. This shortcoming was
resolved to a large extent in\cite{ksk,s1,as} where exact
equations connecting the action functional, effective interaction
and linear response function were derived. But the linear response
function, containing information of the particle-hole and
collective excitations, does not directly present information
about the single particle spectrum.

In this Report,
the self consistent version of the density
functional theory is outlined, which allows to calculate
the ground state and dynamic properties of finite multi-electron
systems starting with the Coulomb interaction. An exact
functional equation for the effective interaction, from which one
can construct the action functional, density functional, the
response functions, and excitation spectra of the considered systems,
is presented. The effective interaction relating the linear response
function of non-interacting particles to the exact linear response
function is of finite radius and density dependent.
We derive  equations describing
single particle excitations of multi-electron systems, using as a
basis the exact functional equations, and show that
single particle spectra do not coincide either with the eigenvalues of
the Kohn-Sham equations or with those of the Hartree-Fock equations.

\section{Exact Equation for the Functional}
Let us briefly outline the equations for the exchange-correlation
functional $E_{xc}[\rho]$ of the ground state energy and
exchange-correlation functional $A_{xc}[\rho]$ of the action
$A[\rho]$ in the case when the system in question is not perturbed
by an external field. In that case an equality holds \cite{ksk,s1}
\beq E_{xc}[\rho]=A_{xc}[\rho]|_{\rho(r,\omega=0)},
\eeq since $A_{xc}$
is also defined in the static densities domain. The
exchange-correlation functional $E_{xc}[\rho]$ is defined by
the total energy functional $E[\rho]$ as
\beq E[\rho]=T_k[\rho] +\frac{1}{2}\int \frac{\rho({\bf
r}_1)\rho({\bf r}_2)} {|{\bf r}_1-{\bf r}_2|}d{\bf r}_1d{\bf
r}_2+E_{xc}[\rho],\eeq where $T_k[\rho]$ is the functional of
the kinetic energy of the non-interacting Kohn-Sham particles.
The atomic system of units
$e=m=\hbar=1$ is used in this paper. The exchange-correlation
functional may be obtained from \cite{ksk}
\beq E_{xc}[\rho]=
-\frac{1}{4\pi}\int \left[ \chi({\bf r}_1,{\bf
r}_2,iw,g')+ 2\pi\rho({\bf r}_1)\delta(w)\delta({\bf r}_1-{\bf
r}_2)\right]
\frac{dwdgd{\bf r}_1d{\bf r}_2}{|{\bf r}_1-{\bf r}_2|}.\eeq
Equation (3) represents the expression for the exchange-correlation
energy of a system \cite{ksk}, expressed via the linear response
function $\chi({\bf r}_1,{\bf r}_2,iw,g')$, with $g'$ being the
coupling constant. For Eq. (3) to describe $A_{xc}[\rho]$ and
$E_{xc}[\rho]$ the only thing we need is the ability to calculate the
functional derivatives of $E_{xc}[\rho]$ with respect to the density.
According to Eq. (3), it means an ability to calculate the functional
derivatives of the linear response function $\chi$ with respect to
the density $\rho({\bf r},\omega)$ which was developed in
\cite{ksk,as,vr}. The linear response function is given by the
integral equation \beq \chi({\bf r}_1,{\bf r}_2,\omega)= \chi_0 ({\bf
r}_1,{\bf r}_2,\omega)+ \int \chi_0 ({\bf r}_1,{\bf r}'_1,\omega)
R({\bf r}\,'_1,{\bf r}\,'_2,\omega) \chi({\bf r}'_2,{\bf r}_2,\omega)
d{\bf r}'_1 d{\bf r}'_2, \eeq with $\chi_0$ being the linear
response function of non-interacting particles, moving
in the single particle time-independent field \cite{ksk,as}. It is
evident that the linear response function $\chi(g)$ tends to the
linear response function of the system in question as $g$ goes to
1. The exact functional equation for $R({\bf r}_1,{\bf
r}_2,\omega,g)$ is \cite{ksk,as} \beq R({\bf
r}_1,{\bf r}_2,\omega,g) =\frac{g}{|{\bf r}_1-{\bf r}_2|}\eeq
$$-\frac{1}{2} \frac{\delta^2}{\delta\rho({\bf r}_1,\omega)
\delta\rho({\bf r}_2,-\omega)} \int\int_0^g\chi({\bf r}_1',{\bf
r}_2',iw,g') \frac{1}{|{\bf r}_1'-{\bf r}_2'|} d{\bf r}_1\,'d{\bf
r}_2\,'\frac{dw}{2\pi}\,dg'.$$ Here $R({\bf r}_1,{\bf
r}_2,\omega,g)$ is the effective interaction depending on the
coupling constant $g$ of the Coulomb interaction. The coupling
constant $g$ in Eq. (5) is in the range $(0-1)$. The
single particle potential $v_{xc}$, being time-independent, is
determined by the relation \cite{ksk,as}, \beq
v_{xc}({\bf r}) =\frac{\delta}
{\delta\rho({\bf r})}E_{xc}[\rho]. \eeq
Here the functional derivative is calculated at $\rho=\rho_0$ with
$\rho_0$ being the equilibrium density. By substituting (3) into
(6), it can be shown that the single particle potential $v_{xc}$
has the proper asymptotic behavior \cite{as,vr}, \beq
v_{xc}(r\to\infty)\to v_{x}(r\to\infty)\to-\frac{1}{r}.\eeq The
potential $v_{xc}$ determines the energies $\varepsilon_i$ and the
wave functions $\phi_i$ \beq
\left(-\frac{\nabla^2}{2}+V_H({\bf r})+V_{ext}({\bf r})
+v_{xc}({\bf r}) \right)\phi_i({\bf r})
=\varepsilon_i\phi_i({\bf r}). \eeq 
These constitute the linear response function
$\chi_0({\bf r}_1,{\bf r}_2,\omega)$ entering Eq. (4)
\beq
\chi_0=\sum_{i,k}
n_i(1-n_k)\phi^*_i({\bf r}_1)\phi_i({\bf r}_2) \phi^*_k({\bf
r}_2)\phi_k({\bf r}_1) \left[\frac{1}{\omega-\omega_{ik}+i\eta}
-\frac{1}{\omega+\omega_{ik}-i\eta}\right] \eeq and the real
density of the system $\rho$, \beq \rho({\bf r})=\sum_i
n_i|\phi_i({\bf r})|^2. \eeq Here $n_i$ are the occupation
numbers, $V_{ext}$ contains all external single particle
potentials of the system, viz. the Coulomb potentials of the
nuclei. $E_H$ is the Hartree energy
\beq E_H=\frac{1}{2}\int\frac{\rho({\bf r}_1)\rho({\bf r}_2)}
{|{\bf r}_1-{\bf r}_2|} d{\bf r}_1d{\bf r}_2,\eeq
with the Hartree potential
$V_H({\bf r})=\delta E_H/\delta\rho({\bf r})$,
and $\omega_{ik}$ is the one-particle excitation energy
$\omega_{ik}=\varepsilon_k-\varepsilon_i$, and $\eta$ is an
infinitesimally small positive number.

\section{The Effective Interaction}
The above equations (2-5) solve the problem
of calculating $E_{xc}$, the ground state energy and the
particle-hole and collective excitation spectra of a system
without resorting to approximations for $E_{xc}$, based on
additional and foreign inputs to the considered problem, such as found in calculations such
as Monte Carlo simulations.
We note,
that using these approximations, one faces difficulties in
constructing the effective interaction of finite radius and the
linear response functions \cite{wks}. On the basis of the
suggested approach, one can solve these problems. For instance, in
the case of a homogeneous electron liquid it is possible to
determine analytically an efficient approximate expression
$R_{RPAE}$ for the effective interaction $R$, which essentially
improves the well-known Random Phase Approximation by
taking into account the exchange interaction
of the electrons properly, thus forming the Random Phase
Approximation with Exchange \cite{s1,as}. The
corresponding expression for $R_{RPAE}$ is \beq
R_{RPAE}(q,g,\rho)=\frac{4\pi g}{q^2}+
\frac{\delta E_x}{\delta\rho}=
\frac{4\pi g}{q^2}+R_E(q,g,\rho), \eeq where
\beq R_E(q,g,\rho)=-\frac{g\pi}{p_F^2}\left[\frac{q^2}{12p_F^2}
\ln\left|1-\frac{4p_F^2}{q^2}\right| -\frac{2p_F}{3q}\ln
\left|\frac{2p_F-q}{2p_F+q}\right|+\frac{1}{3}\right]. \eeq Here
$E_x$ is the exchange energy given by Eq. (3) when $\chi$ is
replaced by $\chi_0$.
The electron density $\rho$ is connected to the Fermi momentum by
the ordinary relation $\rho=p^3_F/3\pi^2$. Having in hand the
effective interaction in $R_{RPAE}(q,g,\rho)$, one can calculate the
correlation energy $\varepsilon^c$ per electron of an electron
gas with the density $r_s$. The dimensionless parameter
$r_s=r_0/a_B$ is usually introduced to characterize the density,
with $r_0$ being the average distance between electrons, and $a_B$
is the Bohr radius.\\

\begin{center}
\begin{tabular}{|r|l|l|l|}  \hline\hline
$r_s$ & $\varepsilon^c_M$ & $\varepsilon^c_{RPA}$ &
$\varepsilon^c_{RPAE}$
\\ \hline\hline
1 & -1.62 & -2.14 & -1.62 \\ \hline 3 & -1.01 & -1.44 & -1.02 \\
\hline 5 & -0.77 & -1.16 & -0.80 \\ \hline 10 & -0.51 & -0.84 &
-0.56 \\ \hline 20 & -0.31 & -0.58 & -0.38 \\ \hline 50 & -0.16 &
-0.35 & -0.22\\  \hline\hline
\end{tabular}
\end{center}\bigskip
In the Table, Monte Carlo
results \cite{mc} $\varepsilon^c_M$ are compared with the
results of the RPA calculation
$\varepsilon^c_{RPA}$, and $\varepsilon^c_{RPAE}$ when the effective interaction $R$ was approximated by
$R_{RPAE}$ \cite{ksk,s1}. The energies per electron are given in eV.
Note that the effective interaction $R_{RPAE}(q,\rho)$ permits the description of
the electron gas correlation energy $\varepsilon^c$ in an
extremely broad range of the variation of the density. At
$r_s=10$ the error is no more than 10\% of the Monte Carlo
calculations, while the result becomes almost exact at $r_s=1$ and is exact when $r_s\to 0$ \cite{ksk,s1}.

\section{Single-Particle Spectrum}
Now let us calculate the single particle energies $\epsilon_i$,
that, generally speaking, do not coincide with the eigenvalues
$\varepsilon_i$ of Eq. (8).
Note that these eigenvalues $\varepsilon_i$ do not
have a physical meaning and cannot be regarded as
the single-particle energies (see e.g.  \cite{wks}). To calculate the
single particle energies one can use the Landau equation \cite{ll}
\beq \frac{\delta E}{\delta n_i}=\epsilon_i.\eeq

In order to illustrate how to calculate the single-particle
energies $\epsilon_{i}$ within the DFT, we choose the simplest
case when the functional $E_{xc}$ is approximated by $E_x$. As we
shall see, the single-particle energies $\epsilon_{i}$  coincide
neither with the eigenvalues calculated
within the Hartree-Fock (HF) method
nor with $\varepsilon_{i}$ of Eq. (8). To proceed, we use a method
developed in \cite{as}. The linear response function $\chi_{0}$ and
density $\rho({\bf r})$, given by Eqs. (9) and (10) respectively,
depend upon the occupation numbers. Thus, one can consider the ground
state energy $E$ as a functional of
the density and the occupation numbers
\begin{equation} E[\rho({\bf r}),n_{i}]=T_{k}[\rho ({\bf
r}),n_{i}]+E_{H}[\rho({\bf r} ),n_{i}]+E_{x}[\rho({\bf r}),n_{i}]+\int
V_{ext}({\bf r})\rho({\bf r})d {\bf r}.  \end{equation} Here $T_{k}$ is
the functional of the kinetic energy of noninteracting particles. As
it follows from Eq. (3), the functional $E_{x}$ is given by
\cite{vr} \beq
E_{x}[\rho]= -\frac{1}{4\pi} \int[\chi_0({\bf r}_1,{\bf r}_2,iw)
+2\pi\rho({\bf r}_1)\delta(w) \delta({\bf r}_1-{\bf r}_2)]
\frac{dwd{\bf r}_1d{\bf r}_2}{|{\bf r}_1-{\bf r}_2|}.\eeq
Upon using Eq. (16), the exact exchange potential
$v_x({\bf r})=\delta E_x/\delta\rho({\bf r})$
of DFT can be calculated explicitly \cite{vr}.
Substituting Eq. (15) into
Eq. (14) and remembering that the single-particle wave functions
$\phi_i$ and eigenvalues $\varepsilon_i$ are given by Eq. (8)
with $v_{xc}({\bf r})=v_{x}({\bf r})$, we see that the single
particle spectrum $\epsilon _{i}$ can be represented by the
expression \begin{equation}
\epsilon_{i}=\varepsilon_{i}-<\phi_{i}|v_{x}|\phi_{i}>+\frac{
\delta E_x}{\delta n_{i}}.
\end{equation} The first and second
terms on the right hand side in Eq. (17) are determined by the
derivative of the functional $T_{k}$ with respect to the
occupation numbers $n_{i}$. To calculate the derivative we
consider an auxiliary system of non-interacting particles in a
field $U({\bf r})$. The ground state energy $E_{0}^{U}$ of this
system is given by the following equation
\begin{equation} E_{0}^{U}=T_{k}+\int
U({\bf r})\rho({\bf r})d{\bf r}.  \end{equation} Varying
$E_{0}^{U}$ with respect to the occupation numbers, one gets the
desired result
\begin{equation}
\varepsilon_{i}=\frac{\delta E_{0}^{U}}{\delta n_{i}}=\frac{\delta
T_{k}}{\delta n_{i}}+<\phi_{i}|U|\phi_{i}>, \end{equation}
provided $U=V_{H}+v_{x}+V_{ext}$.
The third term on the right hand side
of Eq.  (17) is related to the contribution coming from $E_{x}$
defined by Eq. (16). In the considered simplest case when we
approximate the functional $E_{xc}$ by $E_x$, the
 coupling constant $g$ enters $E_x$ as a linear factor.
If we omit the inter-electron
interaction, $g\to 0$, that is, we put $E_x\to 0$,
we directly get from Eq. (17)
$\epsilon_i=\varepsilon_i$ as it must be in the case of a
noninteracting system of electrons. Note that it is not difficult
to include the correlation energy in the simplest local density
approximation
\begin{equation} E_{c}[\rho,n_{i}]=\int\rho({\bf
r})\varepsilon_{c}(\rho({\bf r}))d{\bf r }. \end{equation} Here
the density $\rho({\bf r})$ is given by Eq. (10) and the
correlation potential is defined as
\begin{equation}
V_{c}({\bf r})=\frac{\delta E_{c}[\rho]}{\delta\rho({\bf r})}.
\end{equation}
Varying $E[\rho({\bf r}),n_{i}]$ with respect to the occupation
numbers $ n_{i}$ and after some straightforward calculations, we
obtain the rather simple expression for the single particle
spectrum
\begin{equation}
\epsilon_{i}=\varepsilon_{i}-<\phi_{i}|v_{x}|\phi
_{i}>-\sum_{k}n_{k}\int\left[\frac{\phi_{i}^{\ast }({\bf
r}_{1})\phi_{i}({\bf r}_{2})\phi_{k}^{\ast }({\bf r}_{2})\phi
_{k}({\bf r}_{1})}{|{\bf r}_{1}-{\bf r}_{2}|}\right] d{\bf
r}_{1}d{\bf r}_{2}.  \end{equation} Here $\varepsilon_i$ are
the eigenvalues of Eq. (8) with $v_{xc}=v_x+V_c$.
We employ Eq. (19) and
choose the potential $U$ as
$U=V_{H}+v_{x}+V_{c}+V_{ext}$ to calculate
the derivative $\delta T_{k}/\delta n_{i}$. Approximating the
correlation functional $E_{c}[\rho,n_{i}]$ by Eq. (20), we simplify
the calculations a lot, preserving at the same time the asymptotic
condition, $(v_{x}+V_{c})_{r\rightarrow\infty }\rightarrow -1/r$.
This condition is of crucial importance when calculating the wave
functions and eigenvalues of vacant states within the framework of the DFT
approach \cite{as}. Note, that these functions and eigenvalues
that enter Eq. (22) determine the single particle spectrum
$\epsilon_i$.  This spectrum has to be compared with the
experimental results. The single particle levels $\epsilon_i$, given by
Eq. (22), resemble the eigenvalues $\varepsilon^{HF}_i$ that are
obtained within the HF approximation. If the wave functions $\phi_i$
would be solutions of the HF equations and the correlation potential
$V_{c}({\bf r})$ would be omitted, the energies $\epsilon_i$ would
exactly coincide with the HF eigenvalues $\varepsilon^{HF}_i$. But
this is not the case, since $\phi_i$ are solutions of Eq. (8), and the
energies $\epsilon_i$ do not coincide with either $\varepsilon^{HF}_i$
or with the eigenvalues $\varepsilon_i$ of Eq. (8).
We also anticipate that Eq. (22), when applied to calculations of
many-electron systems such as atoms, clusters
and molecules, will produce
reasonable results for the energy gap separating the occupied and
empty states. In the case of solids, we expect that the energy  gap at
various high-symmetry points in the Brillouin zone of
semiconductors and dielectrics can also be reproduced.

\section{Conclusions}
We have presented the self consistent version of the density
functional theory, which allows to calculate
the ground state and dynamic properties of finite multi-electron
systems. An exact
functional equation for the effective interaction, from which one
can construct the action functional, density functional, the
response functions and excitation spectra of the considered systems,
has been outlined. We have shown that it is possible to calculate the
single particle excitations within the framework of DFT. The
developed equations permit the calculations of the
single particle excitation spectra of any multielectron system such
as atoms, molecules and clusters.  We also anticipate also that these
equations when applied to solids will produce quite reasonable
results for the single particle spectra and energy gap at various
high-symmetry points in the Brillouin zone of semiconductors and
dielectrics. We have related the eigenvalues of the single
particle Kohn-Sham equations to the real single particle spectrum. In
the most straightforward case, when the exchange functional is treated
rigorously while the correlation functional is taken in the local
density approximation, the coupling equations are very simple. The
single particle spectra do not coincide either with the eigenvalues of
the Kohn-Sham equations or with those of the Hartree-Fock equations,
even when the contribution coming from the correlation functional
is omitted.

\section*{Acknowledgments}
The visit of VRS to Clark Atlanta University has been supported by
NSF through a grant to CTSPS. MYaA is grateful to the S.A.
Shonbrunn Research Fund for support of his research. AZM is
supported by US DOE, Division of Chemical Sciences, Office of
Basic Energy Sciences, Office of Energy Research.

\end{document}